\documentclass[apl,preprint]{revtex4}
\usepackage{graphicx}

\begin{document}
\title{Photon- and phonon-assisted tunneling in the three-dimensional charge stability diagram of a triple quantum dot array}
\author{Floris R. Braakman$^{1\ast}$}
\author{Pierre Barthelemy$^{1}$}
\author{Christian Reichl$^{2}$}
\author{Werner Wegscheider$^{2}$}
\author{Lieven M. K. Vandersypen$^{1\ast}$}
\affiliation{1: Kavli Institute of Nanoscience, TU Delft, 2600 GA Delft, The Netherlands}
\affiliation{2: Solid State Physics Laboratory, ETH Z\"{u}rich, 8093 Z\"{u}rich, Switzerland}
\date{\today}

\begin{abstract}
We report both photon- and phonon-assisted tunneling transitions in a linear array of three quantum dots, which can only be understood by considering the full three-dimensionality of the charge stability diagram. Such tunneling transitions potentially contribute to leakage of qubits defined in this system. A detailed understanding of these transitions is important as they become more abundant and complex to analyze as quantum dot arrays are scaled up.
\end{abstract}

\maketitle
Arrays of coupled quantum dots form a promising scalable platform for quantum computation. In this architecture, quantum bits can be defined by spin states~\cite{Loss98,Levy02} or charge states~\cite{Hayashi03}. However, the states spanning the qubit Hilbert subspace are not the only quantum states present in the system. Leakage to states outside the qubit subspace leads to lower fidelities of state initialization, quantum gate operations and read-out~\cite{Devitt07}. To prevent leakage, the qubit states should be as decoupled as possible from other states. For single or double dot systems, this is relatively easy, since other states are few and often far away in energy. However, for larger arrays the Hilbert space is much bigger and the level structure more complicated. It is therefore useful to gain insight into which states are available for leakage in such larger systems and how they can become populated in practice. 

In quantum dot systems, leakage can be triggered for instance by microwave excitation or phonon absorption. Microwaves are used ubiquitously in quantum dot experiments in the context of quantum information for driving single-qubit rotations~\cite{Koppens06,Nowack07,Pioroladriere08}. They are also a useful tool for spectroscopy and characterization~\cite{Wiel03,Oosterkamp98,Petta04} utilizing photon-assisted tunneling (PAT). In PAT, the microwaves induce transitions between tunnel coupled dots that are detuned in energy by an integer multiple of the photon energy. Due to spin-orbit interaction, even states of different spin~\cite{Schreiber11} can be coupled in this way. Microwave excitation can also trigger tunneling transitions between two dots which are separated by an off-resonant third dot, via so-called photon-assisted cotunneling (PACT)~\cite{Braakman13}. 

When applying microwave excitation to drive qubit rotations, PAT may also take place, causing qubit leakage. This limits the microwave power that can be applied when driving Rabi oscillations~\cite{Koppens06,Nowack07,Pioroladriere08}.
Tunneling transitions excited by phonons can produce leakage effects similar to PAT. In particular, charge sensors used for read-out of both charge- and spin-based qubits, have been shown to be a source of phonons~\cite{Aguado00,Granger12}. To mitigate such effects, it is therefore useful to know which photon- or phonon-assisted tunneling transitions are accessible to the system.

Signatures of transitions induced by microwave excitation or by phonons can be found in measurements of charge stability diagrams. In such a measurement, for each dot the voltage on a nearby gate is swept, changing the occupation on that dot. For a double quantum dot such a stability diagram is two-dimensional and relatively fast to measure. For increasing numbers of dots, the stability diagrams grow exponentially, measuring a charge stability diagram of three dots is already quite tedious. In practice, one keeps the measurement time reasonable by taking two-dimensional slices of the stability diagram~\cite{Schroer07}.

In this Letter, we present measurements on a linear triple quantum dot array tuned to the few-electron regime, where we observe tunneling transitions between neighbouring dots induced both by microwave excitation applied to a gate and phonons emitted by a nearby charge sensor. We find that in two-dimensional slices of the charge stability diagram, these transitions can appear in unexpected positions, specifically near the boundary between two charge configurations which the transitions do not connect. We explain these observations by considering the three-dimensional charge stability diagram of the triple quantum dot.

A scanning electron micrograph of a device similar to the one used in the experiment is shown in Figure 1a. Voltage-biased gates (Ti-Au) patterned on the surface of a GaAs/AlGaAs heterostructure electrostatically define the triple quantum dot (indicated by dotted circles), by selectively depleting the two-dimensional electron gas (2DEG) located 85nm below the surface. Another nearby quantum dot (solid circle), created in the same way, functions as a capacitively coupled charge sensor of the triple dot. When positioned on the flank of a Coulomb peak, the conductance through the sensing dot is very sensitive to the number of charges on the triple dot. One of the contacts of the sensing dot is furthermore connected to an LC-circuit, enabling RF reflectometry measurements~\cite{Reilly07} of the conductance. High-frequency lines are connected via bias-tees to  gates LP and RP (Fig. 1a). We apply microwave excitation to gate LP only. The device was cooled inside a dilution refrigerator to a base temperature of $\sim$55mK. All measurements were taken at zero magnetic field.

\begin{figure}[htb]
\centering
\includegraphics[width=0.7\textwidth]{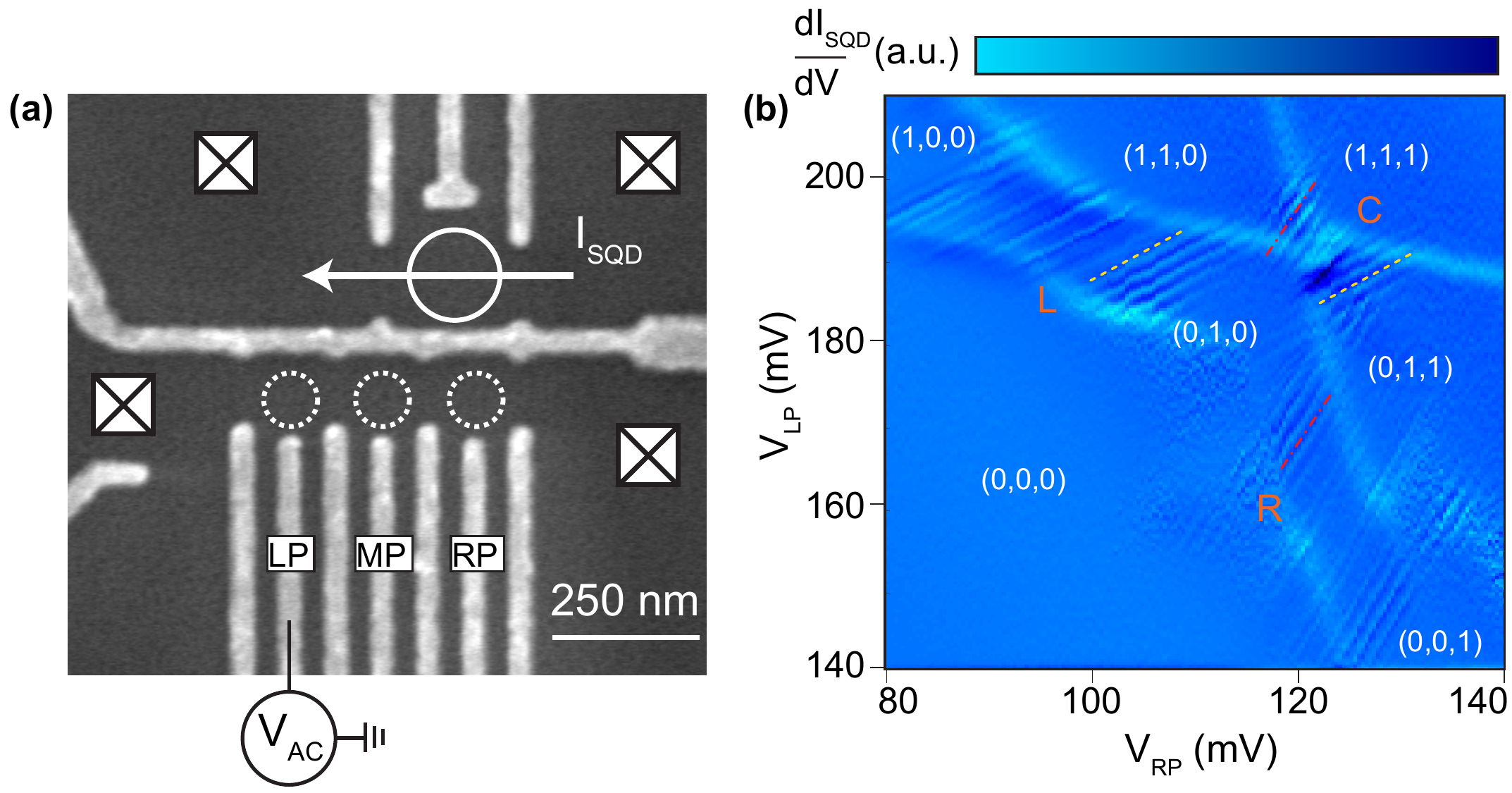}
\caption{(a) SEM image of a sample similar to the one used for the measurements. Dotted circles indicate the linear triple quantum dot, the solid circle indicates a charge sensor dot and squares indicate Fermi reservoirs in the 2DEG which are contacted through Ohmic contacts. The current through the sensor dot (white arrow), as well as its RF reflectance are monitored and used to determine the charge occupancies of the triple dot. (b) Numerical derivative (along $V_{LP}$ axis) of the charge sensor lock-in signal as a function of the voltages on gates LP and RP, while microwaves are applied to gate LP.}
\end{figure}
In the measurements presented in this paper, standard lock-in techniques were used, where a small square wave was applied to gate LP at the lock-in reference frequency of 3412 Hz. The conductance of the sensing dot was monitored through the lock-in amplifier. Whenever microwaves were used, they were chopped at the reference frequency of the lock-in and applied via a bias-tee to gate LP. Figure 1b shows a measurement of the numerical derivative (along vertical axis) of the lock-in signal as a function of the voltage on gates LP and RP ($V_{LP}, V_{RP}$), while microwaves of frequency $\nu$=15 GHz were applied to LP. This maps out a slice of the three-dimensional charge stability diagram of the triple quantum dot. Ground-state charge occupations are indicated as $(n,m,p)$, corresponding to the number of charges on the left, middle and right dot, respectively. Below we denote the lowest energy state with occupation $(n,m,p)$ as $|nmp\rangle$. Regions of stable charge configurations are delineated by charging lines, where electrons can hop between two dots, or between a dot and a reservoir. 

In addition, the presence of the microwave excitation gives rise to many sharp resonances, originating from PAT. The PAT resonances occur at places where the detuning between two dots is equal to an integer times the microwave energy, $h\nu$, with $h$ Planck's constant (see Fig. 2a). At points L and R in Fig. 1b, two such sets of resonances can be discerned, with slopes $S_L$ and $S_R$ respectively. Their slope and location in the charge stability diagram indicate that they correspond to PAT processes between $|100\rangle$ and $|010\rangle$ (point L) and $|001\rangle$ and $|010\rangle$ (point R). As in the case of PAT in a double dot, we can discern pairs of lines with opposite sign (differentiated in Fig. 1b). For instance, near point L we see PAT from $|100\rangle$ to $|010\rangle$, as well as from $|010\rangle$ to $|100\rangle$.

Near point C, two more sets of resonances are present. Note that these sets of lines are located near the $|110\rangle$ to $|011\rangle$ ground-state transition. Since here the $|110\rangle$ and $|011\rangle$ states are nearly degenerate, one might expect transitions between these two states. Even though tunneling then has to take place between non-neighbouring dots, such resonances have recently been reported and referred to as PACT\cite{Braakman13}. They can be distinguished from the resonances under consideration here by their different slope in the stability diagram, corresponding to fixed detuning between left and right dot. In contrast, the right-most lines near point C have slope $S_L$, indicating that the microwaves match the detuning between the left and middle dot. This tells us that the resonances involve tunneling transitions between these two dots. The lines on the left-hand side of point C have slope $S_R$, indicating that they involve tunnel transitions between the right and middle dot.

Using the set of measurements presented in Figs. 2b-g, we will show that the resonances near point C are photon-assisted transitions to the state $|101\rangle$, even when this state is not visible in measurements of the stability diagram. The figures show a series of two-dimensional slices through the charge stability diagram, taken for different values of the voltage on gate MP ($V_{MP}$). Here microwaves of lower power and higher frequency ($\nu$=20 GHz) were used, giving rise to more clearly distinguishable resonances. 

Similar to Fig. 1b, in Fig. 2b, a strong PAT resonance (yellow line) with slope $S_L$ is present in the $(0,1,1)$ region near point C, accompanied by a heavily suppressed parallel line. Additionally, a weak line with slope $S_R$ can be discerned (red line). When stepping $V_{MP}$ to more negative values, we see that the PAT resonances move and change in brightness. The lines with slope $S_L$ move towards the bottom-right of the figures, while the line with slope $S_R$ moves towards the top-left. The suppressed line with slope $S_L$ eventually becomes just as strong as its partner (Fig. 2c, d). This line has coloring opposite to its partner line, indicating that it corresponds to a negative peak in the (non-differentiated) signal. The line with slope $S_R$ also becomes brighter and from Fig. 2e on is accompanied by a partner with opposite coloring. Fig. 2e also shows that a new region (blue shaded) has formed between the two sets of resonances with slopes $S_L$ and $S_R$. By counting charging lines, we see that $(1,0,1)$ is the ground state charge configuration in this region, as indicated in Fig. 2g. 

\begin{figure*}[ht]
\centering
\includegraphics[width=1\textwidth]{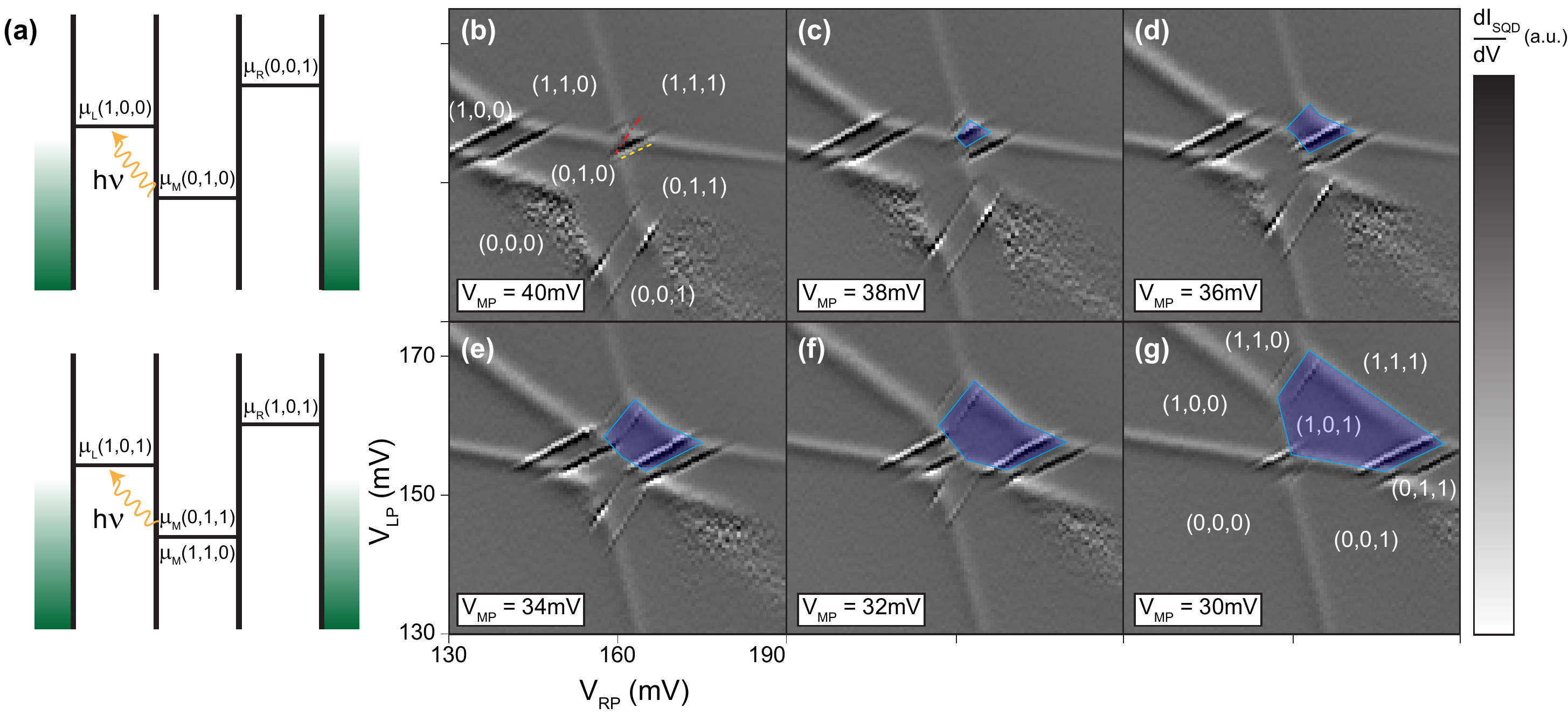}
\caption{(a) Schematic representations of the PAT transitions in terms of the relevant electrochemical potentials of the triple dot, $\mu_i(n,m,p)$, with $i$ standing for the left (L), right (R), or middle (M) dot. The top panel illustrates PAT from $(0,1,0)$ to $(1,0,0)$. The bottom panel shows PAT from $(0,1,1)$ to $(1,0,1)$. Similar PAT transitions involving the middle and right dot can occur for different settings of gate voltages. (b-g) Numerical derivative (along $V_{LP}$ axis) of the charge sensor lock-in signal as a function of the voltages on gates LP and RP, while microwaves are applied to gate LP. The different panels are for different values of $V_{MP}$. The yellow and red lines in the topleft panel indicate PAT transitions between $|110\rangle$ and $|101\rangle$, and $|011\rangle$ and $|101\rangle$, respectively. Indicated in blue is the growing region where $|101\rangle$ is the ground state.}
\end{figure*}

Now it is clear that the resonances with slopes $S_L$ correspond to PAT between $|011\rangle$ and $|101\rangle$, and the resonances with slopes $S_R$ to PAT between $|110\rangle$ and $|101\rangle$. In the two-dimensional slice of Fig. 2b, there is no $(1,0,1)$ ground state charge configuration, but $|101\rangle$ can be reached starting from $|011\rangle$ by photon absorption since it is closeby in the third dimension of the charge stability diagram. However, its partner is strongly suppressed since there is no region where $|101\rangle$ is the ground-state, so there can be no PAT starting out from this state. Thermal activation may populate the $|101\rangle$ state, so this resonance can still be weakly visible. Similarly, this explains why in Fig. 2b only one sideband is visible for PAT between $|110\rangle$ and $|101\rangle$. Note that in Fig. 2g there is again a single PAT resonance present without a partner, this time corresponding to PAT between $|100\rangle$ and $|010\rangle$. In this case there is no region corresponding to $|010\rangle$ being the ground-state.

\begin{figure*}[t]
\centering
\includegraphics[width=0.85\textwidth]{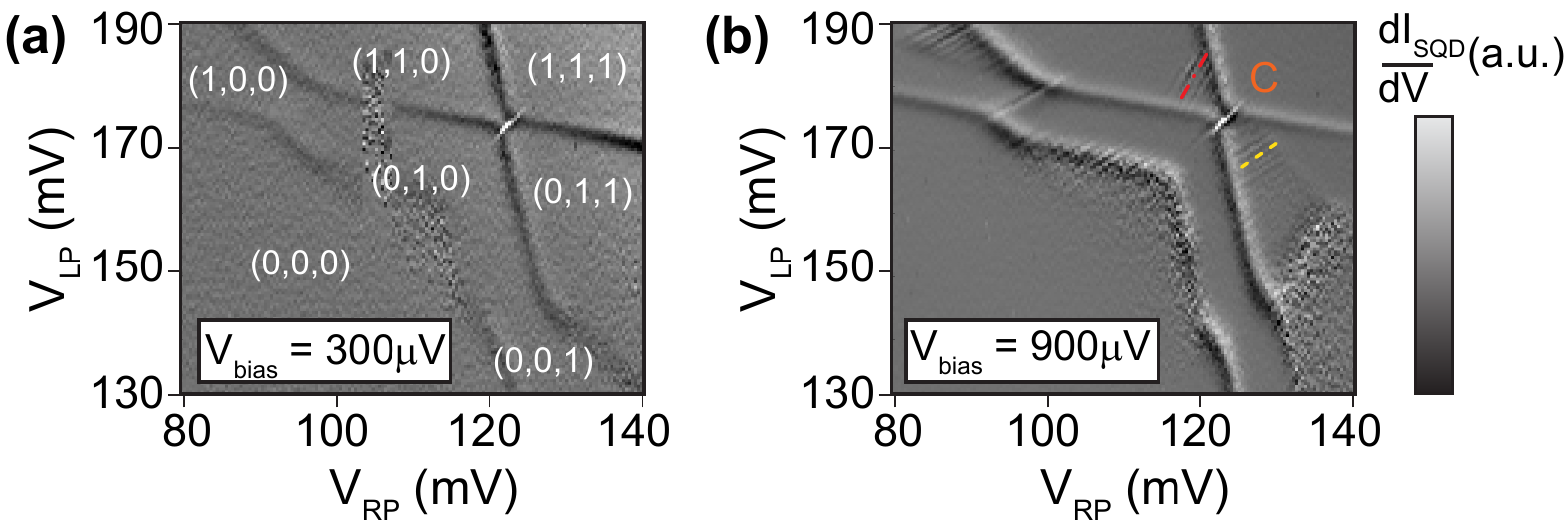}
\caption{(a) Numerical derivative (along $V_{LP}$ axis) of the charge sensor lock-in signal as a function of the voltages on gates LP and RP, with a source-drain bias of the charge sensor dot of $V_{bias}=300 \mu$V. (b) As in (a), but with $V_{bias}=900 \mu$V}
\end{figure*}

Similar transitions can be induced by phonons. Figures 3a, b present charge stability diagrams with no microwaves applied. In Fig. 3a the source-drain bias of the sensing dot $V_{bias}$ was set to 300$\mu$V. In Fig. 3b, $V_{bias}=$900$\mu$V was used and now many additional lines can be discerned. We attribute these lines to transitions induced by the coherent absorption of phonons, as in Granger \textit{et al.}~\cite{Granger12,footnote1}. In the present measurement, inelastic transport of electrons through the sensing dot generates phonons, with a maximum energy set by the source-drain voltage. We focus on the two sets of lines near point C in Fig. 3b. These lines have identical slopes $S_L$ and $S_R$ as the PAT lines discussed above. This indicates that the phonons excite the same transitions as the microwaves in Fig. 2, i.e from $|110\rangle$ to $|101\rangle$, and from $|011\rangle$ to $|101\rangle$. While $(1,0,1)$ does not appear as a ground state in this two-dimensional slice, it is closeby in the third dimension so it can be reached by phonon absorption. Again, the reverse processes, starting from $|101\rangle$ are not visible. 

In summary, we have observed transitions in a triple quantum dot array to a state that is not visible as a ground state in the measured two-dimensional slice of the charge stability diagram. Understanding of these transitions requires consideration of the full three-dimensional charge stability diagram. The transitions can be driven either by microwaves applied to a gate or by phonons generated by a nearby charge sensor. On the one hand, these transitions can lead to leakage of qubit states and should therefore be avoided during qubit manipulation and read-out. On the other hand, the transitions provide the opportunity to efficiently probe the three-dimensionality of the triple dot stability diagram using only two-dimensional measurements. These issues and opportunities become even more important when scaling up to larger quantum dot arrays.

\begin{acknowledgments}
We thank Y. Nazarov, R. N. Schouten, B. v.d. Enden, J. Haanstra, R. Roeleveld, T. Baart, E. Kawakami, P. Scarlino and M. Shafiei for helpful discussions and technical support. This work is supported by the Stichting voor Fundamenteel Onderzoek der Materie (FOM), the Office of the Director of National Intelligence, Intelligence Advanced Research Projects Activity (IARPA), through the U.S. Army Research Office grant W911NF-12-1-0354 and a European Research Council (ERC) Starting Grant.
\end{acknowledgments}


\begin{thebibliography}
\bibitem{footnote1} The regions where such transitions are visible in the charge stability diagram may depend on whether or not metastable charge states can be reached after phonon absorption~\cite{Granger12}.
\end{thebibliography}
\end{document}